\documentstyle[preprint,aps]{revtex}
\begin{document}

\preprint{\vbox{\baselineskip=15pt
\hbox{NSF-ITP-93-84} \hbox{gr-qc/9307009}}}

\title{Late-time behavior of stellar collapse and explosions: \\I. Linearized
perturbations} 

\author{Carsten Gundlach, Richard H. Price and Jorge Pullin}
\address{Department of Physics, University of Utah, Salt Lake City, UT
84112-1195}

\date{9 July, 1993}

\maketitle

\begin{abstract}
We study the power-law tails in the evolution of massless fields
around a fixed background geometry corresponding to a black hole. We
give analytical arguments for their existence at $scri_{+}$, at the
future horizon and at future timelike infinity.  We confirm their
existence with numerical integrations of the curved spacetime wave
equation on the background of a Schwarzschild and a
Reissner-Nordstr\"om black hole.  These results are relevant to
studies of mass inflation and the instability of Cauchy horizons. The
analytic arguments also suggest the behavior of the full nonlinear
dynamics, which we study numerically in a companion paper.
\end{abstract}



\section{Introduction}

In the study of nonspherical gravitational collapse, the late stages
of black hole formation and nonspherical stellar dynamics, certain
simplifications have been traditional. Typically, linearized
perturbation theory has been used on a fixed background, and the
results have been taken to be representative of nonperturbative
collapse.

One of the most basic results \cite{DeBr,RoWi,Ro} about the evolution
of fields on a curved background is that ``sandwich waves'' are not
usually possible.  At late times waves do not cut off sharply but die
off in ``tails.''

In the context of perturbations of spherical objects, like stars or
black holes, arguments have been given \cite{Pr} leading to the
conclusion that the tails have a specific power-law form. The
intention of this paper is to analyze in detail in which regions of
the spacetime this form for the tails holds and under what conditions
they develop.

In section II we start by giving an analytical outline  of the development of
tails in spherically symmetric fixed backgrounds. We present a
somewhat more general and more pedagogical derivation of the results of the
appendix of reference \cite{Pr}. Moreover, we give two new
results:

1) We show that for perturbations around a black hole, power-law tails
develop not only at timelike infinity as was proven in \cite{Pr} but
also at $scri_{+}$ and at the black-hole horizon. This result is of
relevance since the development of tails in these regions is crucial
for the physics of mass inflation \cite{HeKo} and the stability of Cauchy
horizons. At least twice \cite{CuMoPr,Le} in the literature it has
been stated that the power-law tails are ``nonradiative,'' suggesting
that the tails made no appearance at $scri_{+}$ or at the horizon.

2) Generalizing the arguments for the Schwarzschild background, 
we show that power-law tails develop even when no horizon is
present in the background. This would mean, among other things, that
power-law tails should be present in perturbations of stars, or after
the implosion and subsequent explosion of a massless field which does
not result in black hole formation. 

In sections III and IV we confirm the first of these results for the
Schwarzschild background by performing numerical integrations of the
perturbation equations for different initial shapes and different
multipole moments of the field. We also confirm that the results extend
to the case of Reissner-Nordstr\"om black holes; this is the first clear
evidence that power-law tails actually develop for the background of
direct relevance to mass inflation and stability of Cauchy horizons.
Section III contains a brief discussion of our numerical method; section
IV contains our numerical results.

In section V we make some final remarks, especially about possible
implications, both for the behavior of test fields evolving on on a
time-dependent stellar collapse or explosion background, and for the
behavior of a (spherically symmetric) self-gravitating massless field
in a collapse or explosion situation. We develop
this subject in a subsequent paper.

Tails have also been found important in the detailed calculation of
gravitational waveforms from the inspiral collapse of binary systems.
The tail back reaction appear as a correction to the 3/2 Post
Newtonian equations of motions \cite{AKOP}.


\section{Schwarzschild background and beyond}

In this section we examine the evolution of massless perturbation
fields outside a star collapsing to form a black hole. We make the
idealization that the mass loss in the transition from star to black
hole can be neglected. (If there are massless perturbation fields
outside the star at all, they will carry some energy away to null
infinity, but we neglect this compared to the mass of the star.) If
the star is spherically symmetric (again neglecting the
perturbation fields, which need not be spherically symmetric),
spacetime outside the star is Schwarzschild spacetime. We are
therefore dealing with the (massless) wave equation on a region of
Schwarzschild spacetime given by
\begin{equation}
ds^{2}= -\left(1-{2 M \over r}\right) dt^{2}+\left(1-{2 M \over
r}\right)^{-1} dr^{2} +r^{2}(d \theta^{2}+\sin^{2}\theta d\phi^{2})\ ,
\end{equation}
where we have used $c=G=1$ units. We introduce a ``tortoise'' radial
coordinate $r_{*}\equiv r+2M\ln(r-2M)$, advanced time $v\equiv t+r_*$,
and retarded time $u\equiv t-r_*$, in terms of which the line element
becomes
\begin{equation}
ds^{2}= -(1-{2 M \over r}) dudv
+r^{2}(d \theta^{2}+\sin^{2}\theta d\phi^{2})\ .
\end{equation}
We let the star begin its collapse at retarded time $u=u_0$.  The
characteristic $u=u_0$ will be one boundary for our problem.

After the surface has started collapsing, it rapidly approaches the
speed of light. The wordline of the stellar surface is therefore
asymptotic to an ingoing null geodesic $v=v_0$. It is convenient to
consider this $v=v_0$ null geodesic, rather than the stellar surface,
as the left boundary of the problem, and the data on this line for
$u>u_0$ to complete the specification of the problem.  The variation
of $\phi$ on the stellar surface is asymptotically infinitely
redshifted. (See \cite{Pr} for details.) This means that $\phi,_{u}$
will be small (more precisely, exponentially small) at late times and
we make the specific assumption that, after some retarded time
$u=u_{1}$, variations in $\phi$ on $v=v_0$ can be ignored.  This
black hole formation scenario is illustrated in Fig. 1.

Since the background is spherically symmetric, each multipole of a 
perturbation field evolves separately. For a scalar field 
$\phi(t,r,\theta,\varphi)$, for example, we can write $\phi= 
\sum_{l,m}  
\Psi_{m}^{l}(u,r) Y_l^m(\theta,\varphi)/r $
and for each multipole moment we have \begin{equation}
\label{wave}
-4\Psi_{,uv}=V_l(r)\Psi,\qquad\qquad V_l\equiv\left(1-
{2M\over 
r}\right)
\left({l(l+1)\over r^2}+ R(r)\right),
\end{equation}
where $R(r)$ falls off as $r^{-3}$ for large $r$, and where we have 
suppressed the indices $l,m$ on $\Psi$.
The evolutions of scalar, electromagnetic, and gravitational 
perturbations 
are all governed by an equation with the form of (\ref{wave}). Only 
the 
details of $R(r)$ differ from one type of field to another.

The general solution to (\ref{wave}) can be written as a series 
depending 
on two arbitrary functions $F$ and $G$:
\begin{equation}
\label{full}
\Psi=\sum_{p=0}^l A_l^p r^{-p} \left[(G^{(l-p)}(u)+(-1)^p F^{(l-
p)}(v)\
\right]
+\sum_{p=0}^\infty B_l^p(r) \left[(G^{(l-p-1)}(u)+(-1)^p F^{(l-p-
1)}(v)\
\right], 
\end{equation}
where the superindices on $F$ and $G$ indicate the number of 
times the 
function is differentiated; negative superindices are to be 
interpreted as 
integrals. The coefficients $A_l^p$ are dimensionless fixed numbers 
which 
have the same values as in flat spacetime. The coefficient functions 
$B_l^p(r)$, however, vanish for $M\rightarrow0$ and have the 
form 
\begin{equation}
\label{Beq}
B_l^p(r)=a_l^p
r^{-(p+2)}\left[ 1+O(M/r)\right]\ ,
\end{equation}
where the coefficients $a_l^p$ are proportional to $M$. (See 
\cite{Pr} for 
additional details.)

The most interesting characteristic data on $u=u_0$ correspond to two
different cases: a) an initially static perturbation field for $u\le
u_0$, and b) a vanishing perturbation field for $u\le u_0$. The first
case corresponds to the collapse of an initially static star with an
initially static perturbations. The second case corresponds to the
collapse of an initially static star in which the collapse itself
generates the perturbations.

One may also consider a
more general (nonstatic) initial perturbation on $u=u_0$, 
but its generic effects can be modeled by the data on $v=v_0$,
after reflection through the center. We shall see, on
the other hand, that there is a qualitative difference in the tails if
there is a static perturbation present initially.

We break the problem of evolution of the fields into two steps. In the
first step we find $\phi$ on $u=u_1$. This step requires the data on
$u=u_0$, but only the early (small $u$) data on the left boundary. In
this step, then, the type of the left boundary (whether stellar center,
surface of completely collapsing star, etc.) is unimportant.  From the
data on $u=u_1$ we find $\phi$ at late times, and at $scri_{+}$, in a
second step. {\it To leading order in $M$}, the evolution depends on
the spacetime curvature in the first step, but not in the second step.
The validity of this approximation is ultimately justified by
numerical experiment.

We begin with the first step, the scattering in the region
$u_0<u<u_1$. We have taken the variation in $\phi$ on $v=v_0$ to be
negligible after $u_{1}$. This means that, aside from backscattering,
there is no outgoing radiation for $u>u_{1 } $. In (\ref{full}) the
first series represents the non-backscattered waves, so our
assumptions about no late outgoing waves are equivalent to taking
$G(u_1)=0.$ For large $r$ at $u=u_1$ the dominant term in (\ref{full})
is then
\begin{equation}
\Psi(u=u_1,r) =a_l^l r^{-l-2} G^{(-1)}(u_1) \left[  1+O(M/r)\right]\ ,
\end{equation}  
where $G^{(-1)}(u_1)\equiv\int_{u_0}^{u_1} G(u) du$. The characteristic
data, then, is proportional to the integrated initial burst, as well as
to $M$. In the case of an initial static field, with the form
\begin{equation}
\label{statsoln}
\Psi_{\rm static}=\mu r^{-l}\left[  1+O(M/r)\right]\ ,
\end{equation}
the prediction is more definitive. In this case, at $u=u_1$, we have
\begin{equation}
\label{case2}
\Psi=M\mu\left[(2 l+1)/(l+1)\right] r^{-(l+1)}\left[  
1+O(M/r)\right]\ .
\end{equation}
(See again \cite{Pr} for details.)

With this specification of characteristic data on $u=u_1$ we can go on
to consider the subsequent evolution of fields. We will show that the
late time behavior of the fields at constant $r$, and the late-time
behavior at $scri_{+}$, are independent of the details of the background at
small $r$. The evolution is now the same in the spacetime geometry
exterior to a star, a black hole, or to, e. g., an imploding-exploding shell of
scalar field.

We confine our attention to the region $u>u_1, r_*\gg M$.
The leading order effect on the propagation of $\Psi$ is now that of the
``centrifugal term" $l(l+1)/r^2$ in $V_l$.  In this sense we now
approximate spacetime as flat.  The solution for $\Psi$ is then that of
(\ref{full}) with $M=0$,
\begin{equation}
\label{flat}
\Psi(u,v)=\sum_{p=0}^l A_l^p r_*^{-p} \left[(g^{(l-p)}(u)+(-1)^p 
f^{(l-p)}(v)\
\right] \ .
\end{equation}
(To the approximation we are using here, for $r_*\gg M$, we have also $r=r_*$.)
By matching this form to the initial data on $u=u_1$ we find that
\begin{equation}
\label{ftail}
f(v)=F_0/v^P
\end{equation}
where 
\begin{equation}
F_0=(-1)^l 2MG^{(-1)}(u_1), \quad P=2 ,
\end{equation}
if there is no initial static field,and 
\begin{equation}
F_0=(-2)^l 2M\mu\left[ l!/(2 l)! \right], \quad P=1,
\end{equation}
if there is an initial static field, with the form of
(\ref{statsoln}).
(These expressions involve a sum over the $A_l^p$, which are given in
\cite{Pr}.) 

We next take $t\gg r_*$ and we expand $g^{(k)}(u)= g^{(k)}(t) -
g^{(k+1)}(t) 
r_* +\cdots ,$ and similarly for $f(v)$. By reordering terms we 
arrive at 
\begin{equation}
\Psi=\sum_{n=-l}^\infty K_l^n r_*^n 
\Bigl[f^{(l+n)}(t)+(-1)^n g^{(l+n)}(t)\Bigr].
\end{equation}
The coefficients $K_l^n$ 
are constructed from the $A_l^p$ and vanish for $l-n$ even, $-l\le 
n\le 
l$. We define $h\equiv f+(-1)^l g$, and we note that  for $t\gg r^*$
$\Psi$ 
has the form 
\begin{eqnarray}
\label{Taylor}
\Psi=&&K_l^{-l}h(t)r_*^{-l}+K_l^{-l+2}h''(t)r_*^{-l+2}+...\\
&&+K_l^lh^{(2l)}(t)r_*^l+K_l^{2l+1}\Big[2f^{(2l+1)}(t)-
h^{(2l+1)}(t)\Big]\,
r_*^{l+1}+...\nonumber
\end{eqnarray}
We know that at late times $f(v)$ has the power-law form given in 
(\ref{ftail}); we make the {\em ansatz} that $g(u)$ also falls off as a 
power law, so that we may write $h(t)=H_0/t^N$, where $H_0$ is a 
constant.

We note that as $t\rightarrow\infty$ the term $h(t)r_*^{-l}$ will
dominate if $N<P+2l+1$ and the term $2f^{(2l+1)}(t)r_*^{l+1}$ will
dominate if $N>P+2l+1$. In either case only one of the constants $H_0$
or $F_0$ has an influence on the solution, and in fact the constant is
simply an overall scaling of the solution. If either $N<P+2l+1$ or
$N>P+2l+1$, therefore, the form of the solution (aside from overall
scaling) is fixed for $r_*\gg M$, its continuation to smaller values
of $r_*$ is fixed, and we cannot satisfy boundary conditions for small
$r$ (e.g., regularity at $r=0$ for a stellar model or, as
$r_*\to-\infty$, at the horizon of a black hole). In order to satisfy a
small-$r$ boundary condition we must have $N=P+2l+1$ and hence
\begin{equation}
\label{same}
f(t)\simeq F_0 t^{-P},  \quad  g(t)\sim (-1)^{(l+1)} F_0 t^{-P},
\end{equation}\begin{equation}
h(t)\equiv f(t)+(-1)^l g(t)\sim 1/t^{P+2l+1}.
\end{equation}

For $M\ll r_*\ll t$ the form of the fields is particularly
simple. In that case we have
\begin{equation}
\label{larger}
\Psi=K_l^{2l+l} 2 f^{(2l+1)}(t) r_*^{l+1}
=-2 K_l^{2l+l} F_0 (P+2l)! t^{-(P+2l+1)} r_*^{l+1} \ ,
\end{equation}
and $\Psi$ therefore falls off as $1/t^{2l+2}$
(initial static field) or $1/t^{2l+3}$ 
(no initial static field) at timelike infinity $i_+$. It 
follows from (\ref{same})
and (\ref{flat}) that
at $scri_{+}$ (i. e., at $v\rightarrow\infty$) we have 
\begin{equation}
\Psi(v\to\infty,u)\simeq A_l^0 g^{(l)}(u)\simeq -A_l^0 F_0(l+P-1)! u^{-
P-1}\ .
\end{equation}
At null infinity $scri_{+}$ the fields therefore fall off as $u^{-l-1}$ 
(static initial field) or as 
$u^{-l-2}$ (no static initial field).

We keep in mind that the above analysis did not depend on the
small-$r$ details of the problem, and we go on to consider the
specific case of a black hole.  (It does not matter if it is eternal
or formed in collapse, only that there is an internal ``infinity", i.
e. an event horizon.)  As $r_*\rightarrow- \infty$ the curvature
potential $V_l$ in (\ref{wave}) is negligibly small and the solution
to (\ref{wave}) can be written as $\Psi=\alpha(u)+\gamma(v)$.  The
nature of the characteristic data at $v=v_0$ requires that $\alpha(u)$
be a constant (aside from exponentially small variation) and with no
loss of generality we choose it to be zero. For $|r_*|\ll t$ we can
then expand the solution, for large $u$ and $r_*\ll-M$ as
\begin{equation} \label{lateH}
\Psi=\gamma(v)=\gamma(t)+\gamma'(t)r_*+\frac{1}{2} \gamma''(t)
r_*^2+\cdots\ .
\end{equation}

To join this solution, at $r_*\ll-M$ to our previous solution in 
(\ref{larger}) 
for $r_*\gg M$, we make one further assumption. We assume that 
when 
$t\gg|r_*|$ then, for the whole range of $r_*$ (from $r_*\ll-M$ to  
$r_*\gg M$), the solution has the form $\Psi\approx\Psi_{\rm 
finstat}(r)/t^{P+2l+1}$, where $\Psi_{\rm finstat}(r)$ is a $t$-
independent 
solution of (\ref{wave}). This is clearly the case in the region 
$t\gg r_*\gg M$.
With this assumption we can match the solution in (\ref{lateH}) 
at $r_*\ll-M$ and that in (\ref{larger}) for
$r_*\gg M$, and conclude that $\gamma(t)=\Gamma_0 t^{-P-2l-
1}$, and 
therefore that the late time behavior at the horizon is 
\begin{equation}
\label{utoinf}
\Psi(u\to\infty,v)=\gamma(v)=\Gamma_0 v^{-P-2l-1}\ .
\end{equation}
The constant $\Gamma_0$ is determined by the condition that there 
is a 
static solution $\Psi_{\rm finstat}$ such that 
\begin{equation}
\label{horlimit}
\lim_{r_*\rightarrow -\infty}\Psi_{\rm finstat}(r)
=\Gamma_0,
\end{equation}\begin{equation}
\label{infrlim}
\lim_{r_*\rightarrow \infty}\Psi_{\rm finstat}(r)
=-2K_l^{2l+1} F_0(P+2l)! r^{l+1}.
\end{equation}
The coefficient $\Gamma_0$, and therefore the behavior at the 
horizon, 
like all other aspects of the late time behavior, is fixed by the initial 
backscattering.

In the above analysis, we have seen that the backscattering of the
initially outgoing waves, and the subsequent evolution in time,
produces the power-law tails at $scri_{+}$ and at future null infinity
$i_ +$. The small-$r$ nature of the background does not enter (except,
of course, in the discussion of the tails at the horizon).  This means
that the same power-law tails should develop at $scri_{+}$ and at
$i_+$ in models other than black hole models. We might consider as
backgrounds: incompletely collapsing stars, static stars, imploding
and exploding shells, and so forth. These different models would have
different small-$r$ boundary conditions. All that is important to tail
formation is that the source of the perturbations is sharply cut off
as happens, due to the infinite redshift, in the black hole collapse
case. We might consider, as an example, a quadrupole deformation of a
nonrotating neutron star. A dynamical process might change the
quadrupole perturbation from one static value to another (nonzero)
static value. In this case, clearly, there cannot be a power-law fall
off of the perturbation in time; the perturbation does not fall off at
late times. On the other hand there might be a phase change in which
the stellar crust loses the ability to support shear stresses
responsible for the quadrupole moment, and the star might quickly
become spherical. In this case, our analysis predicts that at late
times the exterior quadrupole perturbation will fall off as $1/t^{6}$.

The only specific detail of our analysis that must be modified for
non-hole models is the nature of the small-$r$ solution. In
particular, the horizon condition in (\ref{lateH}), (\ref{utoinf}),
and (\ref{horlimit}) must be replaced by the appropriate small-$r$
condition, and the static solution $\Psi_{\rm finstat}(r)$ is no
longer the solution well behaved at $r\rightarrow 2M$, but rather the
solution satisfying the appropriate small-$r$ condition (e.g.,
regularity at $r=0$). The result in (\ref{infrlim}) remains unchanged.

The analysis of perturbations on a Reissner-Nordstr\"om (RN) 
background, of mass $M$ and charge $Q$, is very similar to that 
given 
above for the Schwarzschild background. For scalar perturbations 
or for 
all radiative degrees of freedom of the mixed electromagnetic-
gravitational perturbations, the field equations can be put in the 
form of 
(\ref{wave}), but with a potential of the form 
\begin{equation}
\label{RN}
V_l(r)=\frac{(r-r_+)(r-r_-)}{r^2}\left[ 
\frac{l(l+1)}{r^2}+R(r)\right]\ ,
\end{equation}
where $r_+, r_-$ are the radii of the outer and inner horizons
\begin{equation}
r_\pm=2M\left(1\pm\sqrt{1-(Q/M)^2}\right)\ ,
\end{equation}
and where $R(r)$ falls off as $r^{-3}$. The tortoise coordinate $r_*$ 
for the 
RN background is the solution of 
\begin{equation}
\frac{dr_*}{dr} =\frac{r^2}{(r-r_+)(r-r_-)} 
\end{equation}
for $r_+<r<\infty$.

A review of the analysis of late time behavior for the Schwarzschild
background confirms that almost all of the arguments depend on the
general form of $V_l(r)$ at large $r$ and the exponentially sharp fall
off of $V_l(r)$ at the horizon. In these features there seems to be no
difference between the Schwarzschild and the RN problems. The
difference in the relationship of $r$ and $r_*$ in the two cases must
be carefully considered, however; it is this relationship [not the
details of $R(r)$] that determines the initial backscattering and
therefore the behavior of the late time tails. It turns out that there
is no difference (for the tails) between the Schwarzschild and the RN
cases. When the expression in (4) is substituted in (3) (with the RN
form of $u,v$ and $V_l$) the result in (5) again emerges, and it is
this result that determines the initial backscattering. The
differences between the RN and Schwarzschild cases enter into the
determination of $B_l^p(r)$ only to higher order in $r^{-1}$.


\section{Numerical method}

It is straightforward to integrate equation (\ref{wave}) on a $uv$ null
grid. The two-dimensional wave equation
$-4\Psi_{,uv}=V_{l}(r) \Psi$ is most simply
discretized as
\begin{equation}
\Psi_N=\Psi_E+\Psi_W-\Psi_S-\Delta u\,\Delta v\, 
V_l\left({v_N+v_W-u_N-u_E\over4}\right){\Psi_W+\Psi_E\over8} +O(h^4).
\end{equation}
Here the points N, S, E and W form a null rectangle with relative
positions indicated by the points of the compass, and $h$ is an
overall grid scale factor, so that
\begin{equation}
\Delta u=u_N-u_E=u_W-u_S=O(h),\qquad \Delta v=v_N-v_W=v_E-v_S=O(h).
\end{equation}
Starting from null data on $u=u_0$ and $v=v_0$, integration proceeds 
to the northeast (increasing $v$) on one $u=const.$ line after another.

The error estimate for $\Psi_N$ follows directly if one assumes that
the exact solution $\Psi(u,v)$ has a Taylor expansion in the given
null rectangle.  On a grid of fixed total size in $u$ and $v$ there
are $O(h^{-2})$ grid points, so that the total error on the far end of
the grid from the null data should scale as $h^2$ when the grid size
is scaled by $h$. We have done numerical tests which have confirmed this
$O(h^2)$ convergence.

The only difficulty in developing a code for this problem was the
calculation of $r$ from $r_*=(v-u)/2$. We have done it by iteration of
the defining equation, in the form $r=r_*-\ln(r-1)$ for large $r$, and
in the form $r_1=\exp(r_*-1-r_1)$ for $r$ close to $1$, that is,  close to
the horizon, where $r_1\equiv r-1$. We have used a grid of constant
$\Delta u=\Delta v$. Because of the scale-invariance of the problem we
have set $2M=1$.

There is a small difference between the null data on $u=u_0$ and $v=v_0$
which we posed in the previous, analytical, section and our numerical
work. For the numerical work, we posed {\it constant} data on $v=v_0$,
and either generic (here Gaussian) or static data on $u=u_0$. The
justification for this is the following. As we consider a linear wave
equation, we can examine the evolution of data on $v=v_0$ and of data on
$u=u_0$ separately. If we want to consider generic data on $v=v_0$, i.
e. data which are localized, we can reflect them back at the scattering
potential to get generic (although different data) on some $u=\tilde
u_0$. So if we are not very interested in the exact shape of our data,
putting them on $u=u_0$ is sufficient.


\section{Numerical results}

We now report on the results of our numerical simulations of test
fields
on a Schwarzschild or RN background. Without loss of generality we
have set $2M=1$. As we are dealing with a linear wave equation, the
overall amplitude of our initial data can also be chosen arbitrarily
and is physically irrelevant.
As our initial data null surfaces we chose $v=0$ for $u\ge0$ and $u=0$
for $v\ge 0$, which meet at the point $t=0$, $r_*=0$. (The background is
of course $t$-independent.)

In a first series of numerical experiments, we chose $V_{l}(r)$
appropriate for a massless minimally coupled scalar field on a
Schwarzschild background for different multipole
indices $l$.  It is of the form (\ref{wave}), with $R(r)=2M/r^3$.
As null data we used a Gaussian of width $3$ centered
at $v=10$ on $u=0$. $\Psi$ is chosen to be constant on the null
boundary $v=0$, with the constant determined by $\Psi(u=0,v=0)$.
This is a simple approximation of the idea that the field is anchored
in a star whose surface is rapidly redshifting, as explained in
section II.  We extended the grid from $v=0$ and $u=0$ to $u=400$ and
$v=400$, with a typical value of $\Delta u=\Delta v=0.1$.

We examined the value of $\Psi$ as a function of $t$ on three lines,
namely $r_*=10$, $u=400$ and $v=400$. We took these lines as finite
approximations of the future timelike infinity $i_+$, of the future
horizon ${\cal H}_+$ and future null infinity $scri_+$. Log-log plots
of these three ``tails" for $l=0$ are shown in Fig. 2.  The predicted
power-law behaviors are apparent for the case of an initial static
field. On ${\cal H}_+$ and $i_{+}$ the field falls off as $t^{-3}$; on
$scri_+$ the fall-off is as $t^{-2}$. The bend at the end of the
$scri_+$ line is not surprising. It represents null infinity only for
$v\gg u$, which is no longer the case at that end of the line. The
large wiggles at the left are a remnant of quasinormal ringing.

Quasinormal ringing is shown more clearly in the linear plot, for
$l=1$ scalar perturbations, shown in Fig. 3.  The agreement with
the theoretically calculated frequency \cite{Le} is good. From the
plot we read off values of $0.56$ and $-0.19$ for the real and
imaginary parts of the frequency, for a scalar field with $l=1$, for
$2M=1$. The predicted values \cite{Le} are $0.58587$ and $-0.19532$ for the
least damped modes.

In Fig. 4 we show the effect of the multipole index  $l$ on the
power law of the tail at $i_+$, again for a scalar field and Gaussian
null data. The agreement of the power laws with the prediction is
striking.

In a second series of numerical experiments we looked at the potential
$V_l$ appropriate for an electromagnetic field on a
Schwarzschild background, which is (\ref{wave}) with $R(r)=0$.
We took null data corresponding to the
initial presence of a static $l$-pole moment. 
(These are given in power-series form in \cite{CuMoPr}, and we
numerically
evaluated the first few lowest orders in $1/r$.)
The tails at constant
$r$
for $l=1$ and
$l=2$ are shown in figure 5. (There is no electromagnetic monopole
field.) It was shown in \cite{Pr} how electromagnetic and
spin-2 fields can be encoded in a scalar field. The potential $V_l$
depends on the spin as well as on $l$, but the terms by which it
differs for different spins are essentially Riemann tensor components
and therefore of order $2M/r^3$. Terms of this order have been
neglected in our analytic derivation, and the surprising prediction is
therefore that the power law of the tails is independent of the spin
of the testfield.

In a third series of experiments we examined a scalar test field on 
Reissner-Nordstr\"om backgrounds. The potential for a scalar field is now
of the form (\ref{RN}), with $R(r)=2Mr^{-3}-2Q^2r^{-4}$
As explained in section III one expects
the tails to be independent of the charge of the black hole. Figure 6
shows that this is in fact so, with the example of the tails at the (outer)
horizon of an $l=0$ test field on Reissner-Nordstr\"om backgrounds with
charge/mass ratios of $Q^2/M^2=0.1$ and $0.9$ respectively. This figure
constitutes also direct evidence for the existence of power-law tails
on the (outer) horizon for a generic perturbations, thus underpinning
a central condition for the mass inflation scenario.


\section{Conclusions}

When the dust of an approximately spherical collapse has settled, and
spacetime inside the future lightcone of the collapse has approached
Schwarzschild, Minkowski, or a stellar interior spacetime, any
massless fields that were present in the collapse still show ``tails"
that linger, decaying only as a power of time. In particular the
$l$-th multipole moment of a massless test field decays like
$t^{-(2l+P+1)}$ at fixed radius at large times, with $P=1$ if there is
an $l$-pole moment present before the collapse and $P=2$ otherwise.  But
power-law tails are also present on $scri_{+}$, where they decay like
$u^{-(l+P)}$, and if a black hole has formed, on the horizon, where
they decay like $v^{-(2l+P+1)}$. The amplitude of the tails can be
calculated as a function of the initial multipole moment, and in its
absence, as an integral over the radiation going out to infinity
during the collapse.

Neither this amplitude, nor the exponent of the power law, depend upon
the spin of the massless field in question, nor, if the black hole is
charged, on its charge. We have shown the origin of these features in
an analysis which is based on \cite{Pr}.  The spin- and
charge-dependent parts of the effective radial potential of a massless
test field on a Reissner-Nordstr\"om background are of an order that can
be neglected in this analysis.

We have checked numerically that the tails are indeed present on null
infinity and the horizon, and are independent of the spin of the field
and the black hole charge.  The fact that power-law tails are present
on the outer horizon of a Reissner-Nordstr\"om black hole after a
generic collapse situation is of crucial importance to the mass
inflation scenario. It has, to our knowledge, never been demonstrated
explicitly.

Finally, one decisive step of our analysis was a regularity condition,
either on the horizon when a black hole formed in the collapse, or
else at the center of spherical symmetry.  This argument generalizes
to any kind of boundary condition posed at small radius, and strongly
suggests that perturbations of massless fields outside any spherical
background object should also have power-law tails with the powers
given above. In a subsequent paper we report results for a closely
related {\em nonlinear} problem: the implosion of a shell of scalar
field. 

\thanks

We wish to thank Josh Goldberg and Fritz Rohrlich for pointing out
early references.  This work was supported in part by grant NSF
PHY92-07225 and by research funds of the University of Utah.
J.P. acknowledges hospitality and support from the Institute for
Theoretical Physics at UCSB and the National Science Foundation grant
PHY89-04035.

\figure{Fig. 1: Spacetime of a collapsing spherically symmetric
star. Outside the star the metric is Schwarzschild. Coordinates $t$,
$r_*$, $u$ and $v$ are indicated.}

\figure{Fig. 2: Log-log plots of a spherically symmetric 
scalar test field on Schwarzschild
spacetime. The initial data were Gaussian.
$\phi(r=10,t)$ represents future timelike infinity.
$\phi(u=400,v)$ represents the future horizon. $\phi(v=400,u)$
represents future null infinity. The power-law exponents are $-3.08$ on
$i_+$ and $-3.02$ the horizon, and $-2.11$ on $scri_{+}$.}

\figure{Fig. 3: $\phi(r=10,t)$ plotted for a scalar testfield with
angular dependence $l=2$ on Schwarzschild spacetime, showing
quasinormal ringing. The initial data are Gaussian.}

\figure{Fig. 4: Log-log plots of $\phi(r=10,t)$ for  scalar
testfields with angular dependence $l=0,1,2$, showing power-law
exponents $-3.03$. $-4.99$, and $-6.94$. Gaussian initial data.}

\figure{Fig. 5: Electromagnetic test field, with angular
dependence $l=1,2,3$ on Schwarzschild background. Plotted is the
field at $u=400$ versus $v$, representing the horizon.
Static-static
initial data. The power-law exponents are $-4.95$ and $-5.93$ and
$-8.63$.}

\figure{Fig. 6: Scalar test field on RN background, shown on
"horizon". Gaussian initial data with $l=0$. Shown are charge/mass
ratios $Q^2/M^2=0.1$ and $0.9$. The power laws are $-3.18$ and
$-3.17$.}

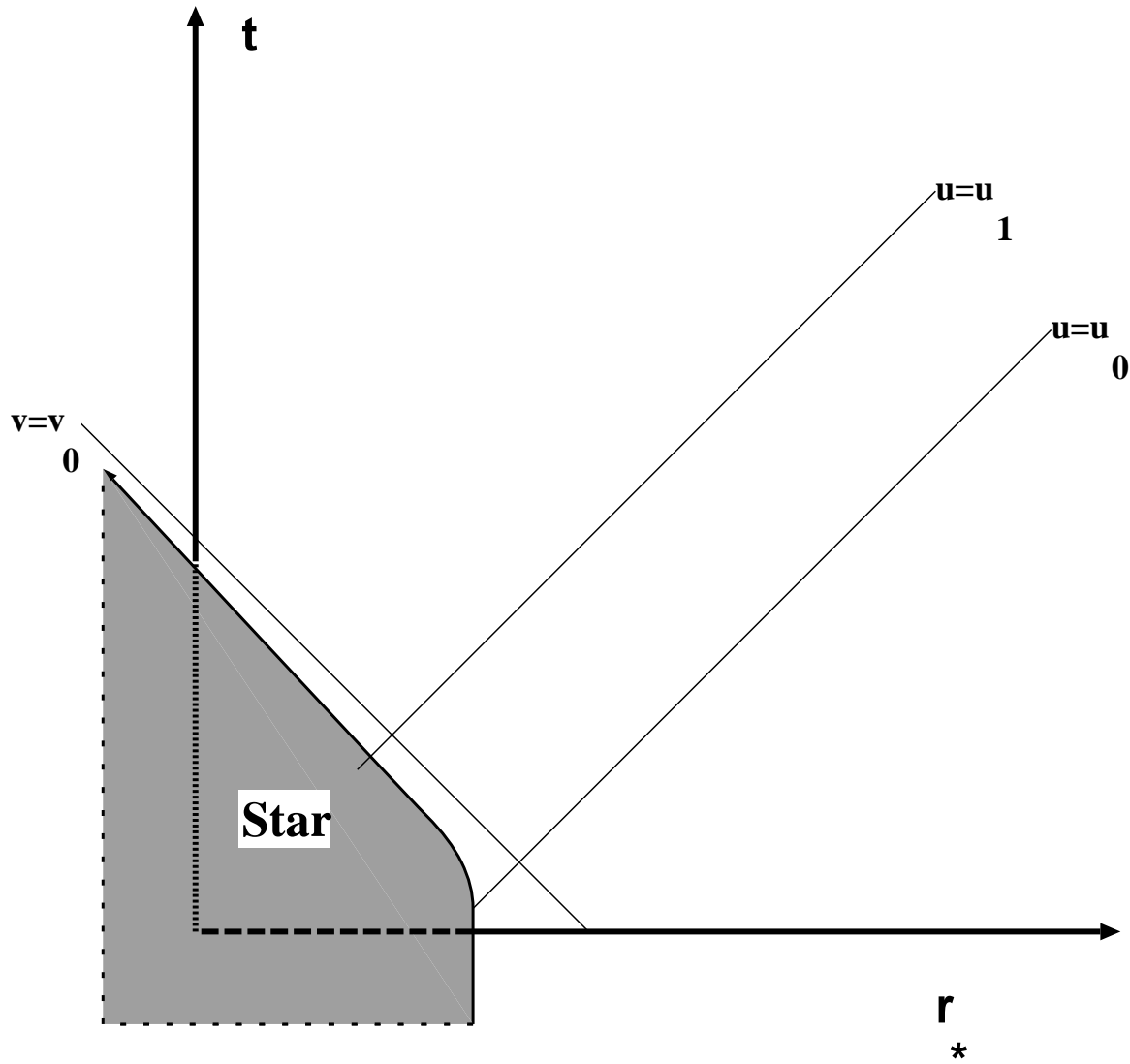
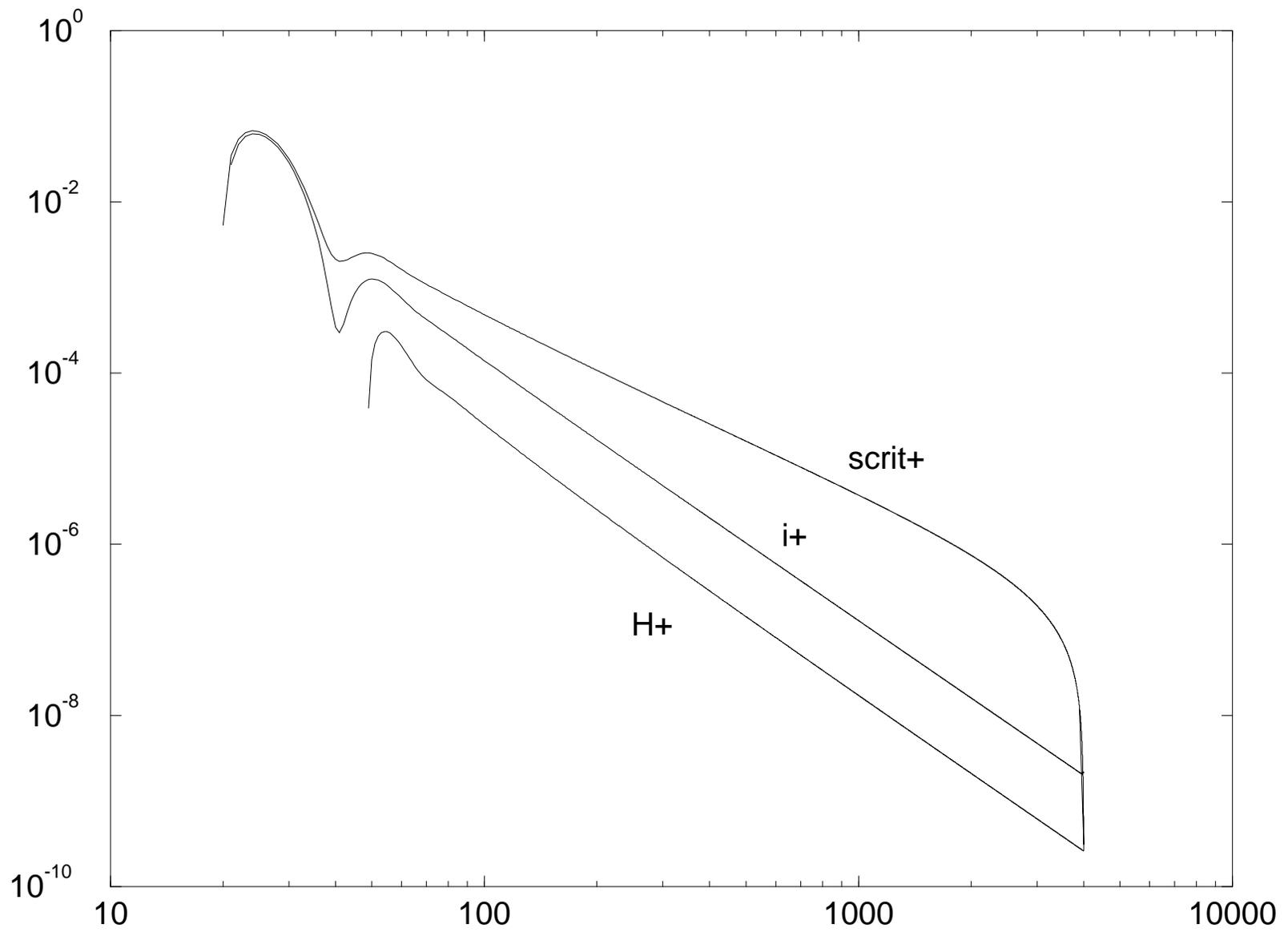
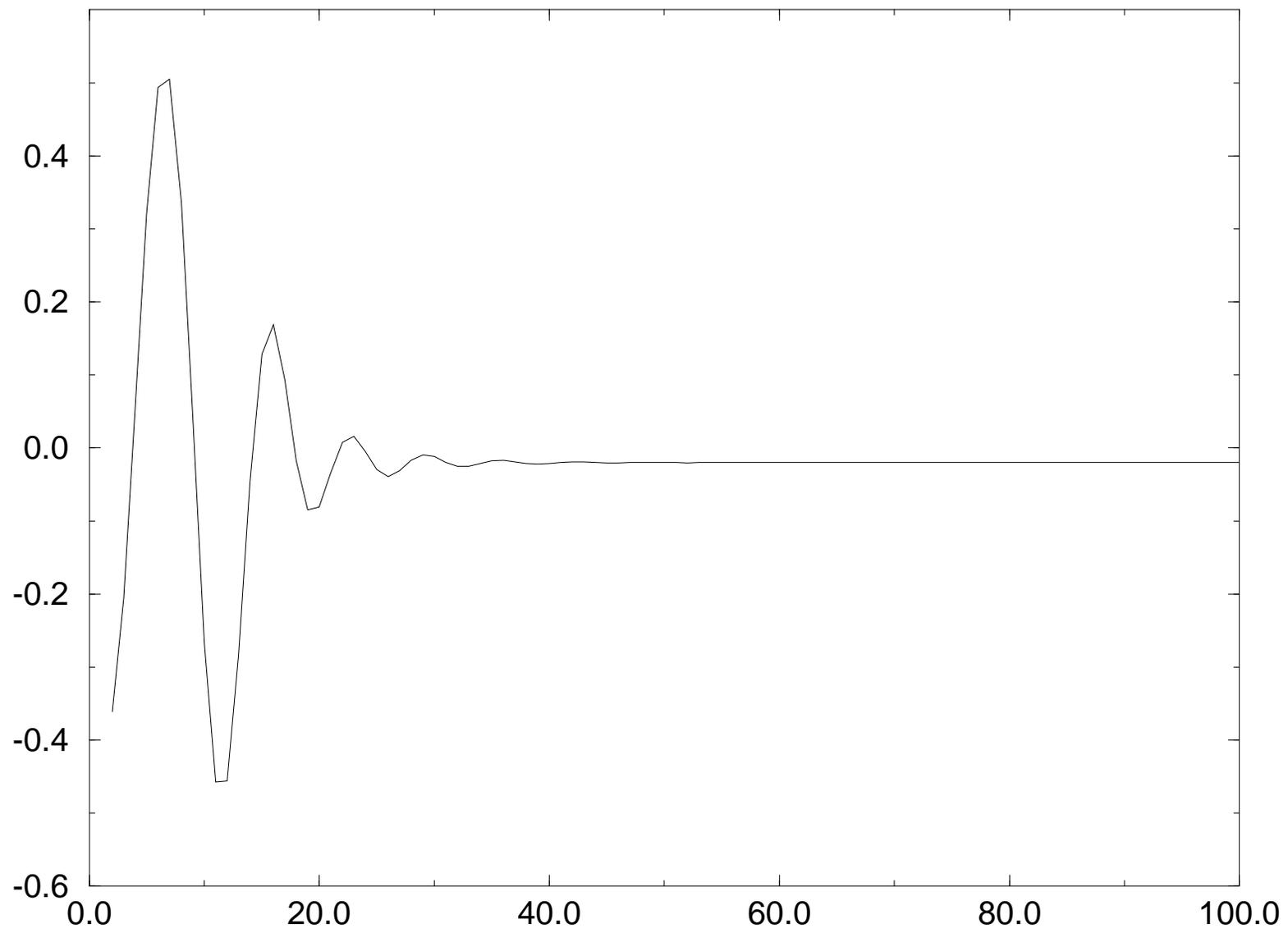
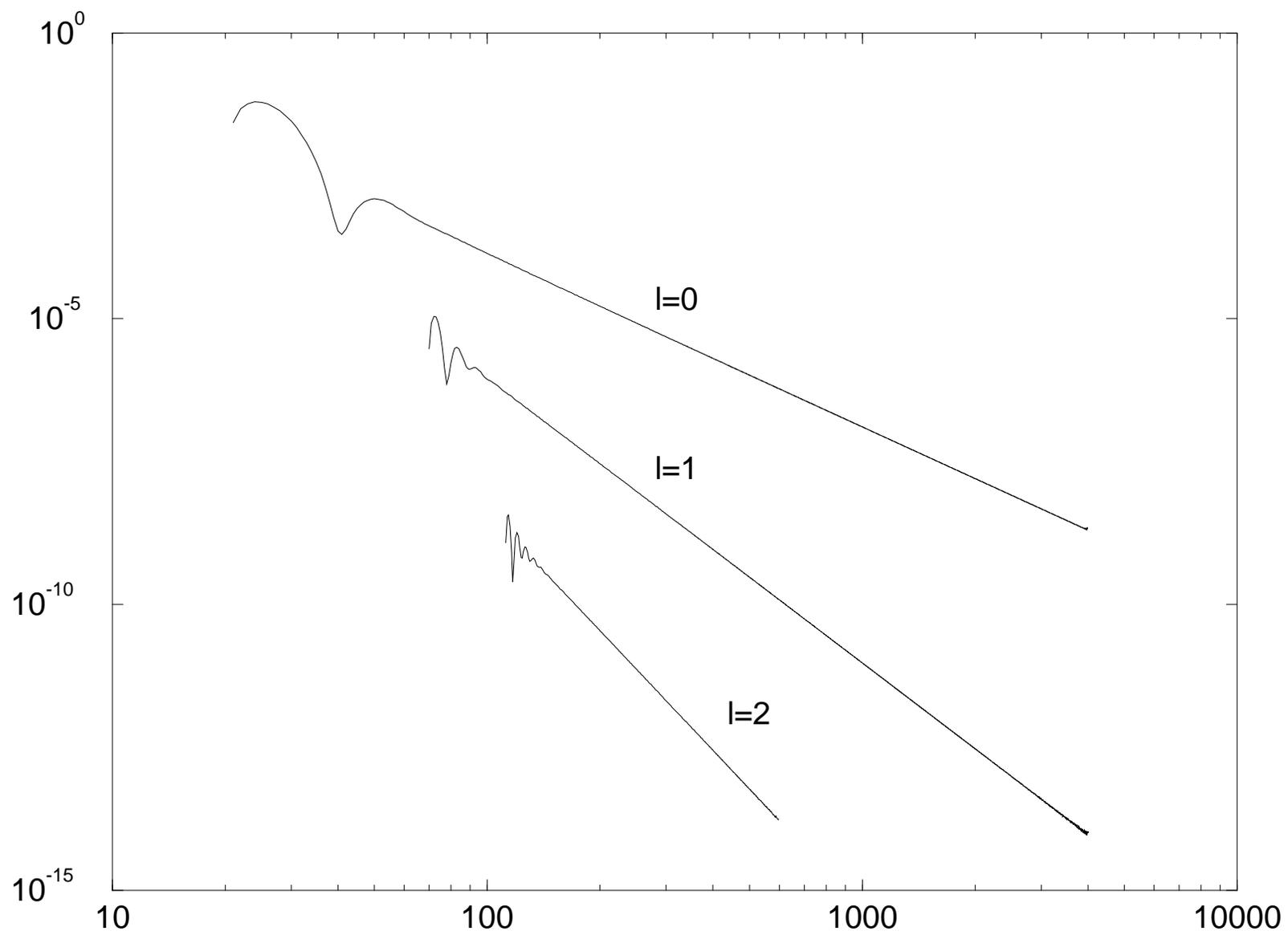
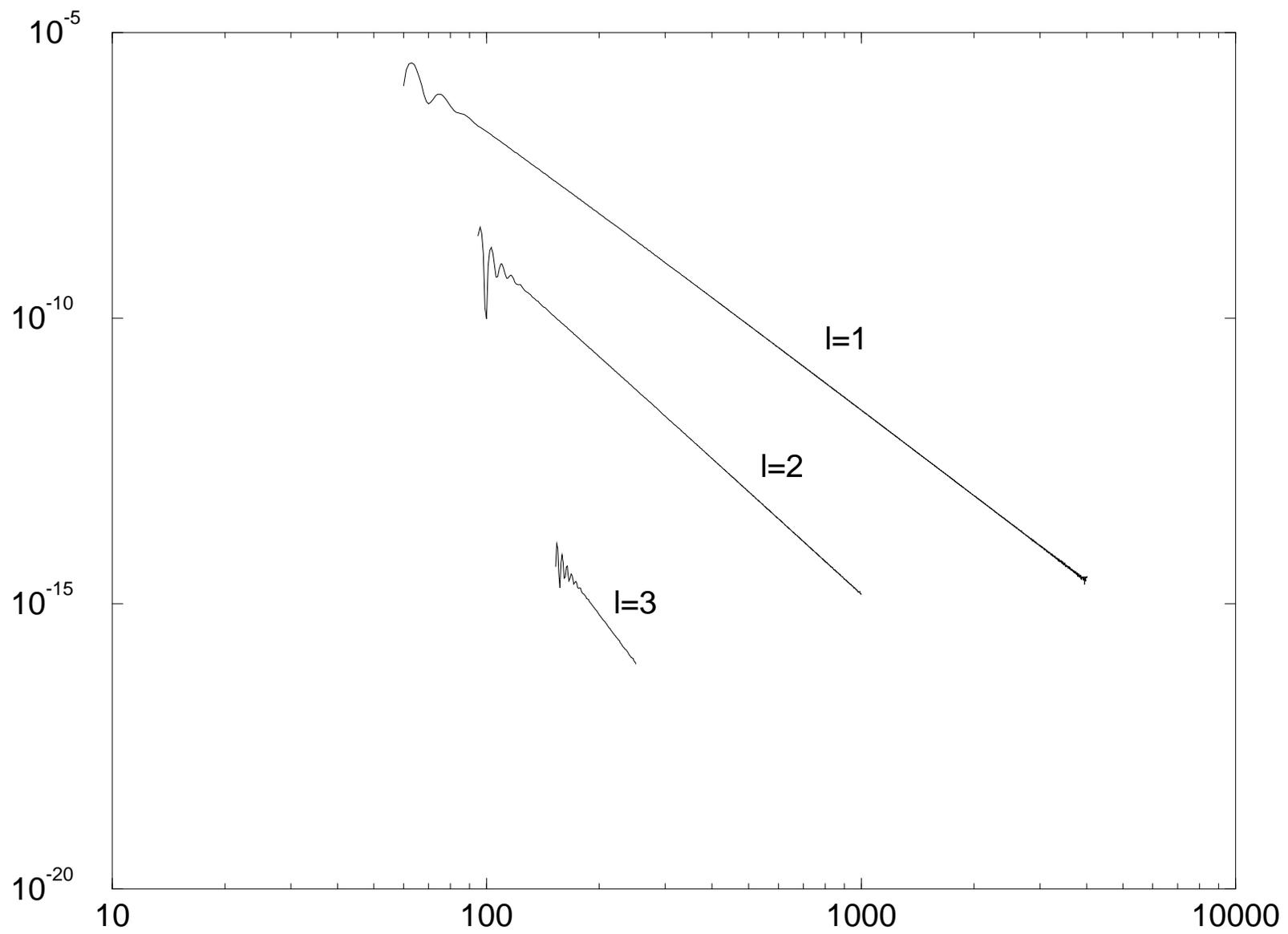
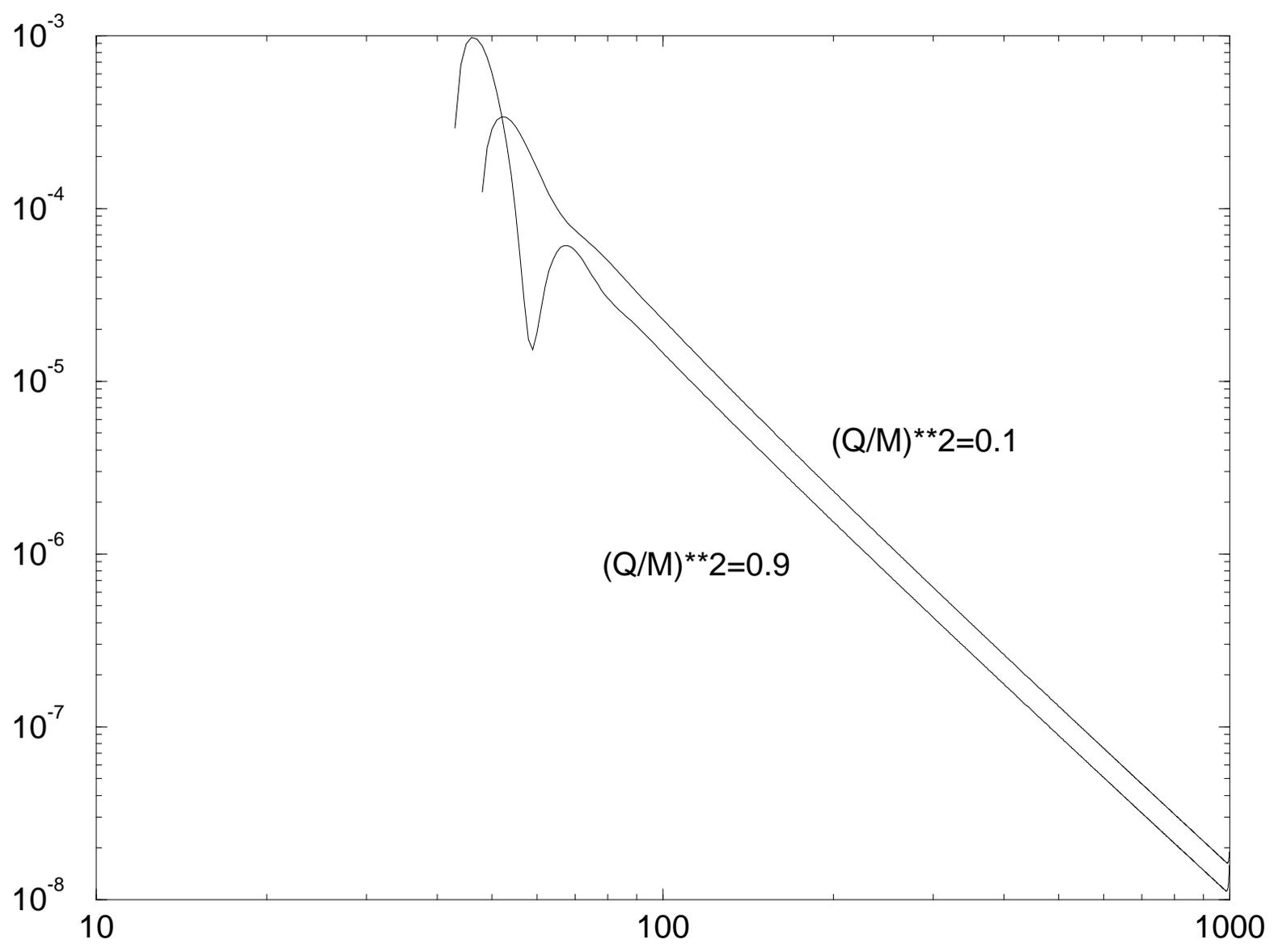
\end{document}